\documentstyle[preprint,aps]{revtex}

\def\addcontentsline#1#2#3{\relax}
\begin{document}
\title{
Correlations in one-dimensional disordered electronic systems\\
with interaction
}
\author{Masanori Yamanaka$^{1, *}$ and Mahito Kohmoto$^{2, **}$}
\address{
$^1$Department of Applied Physics, University of Tokyo,
7-3-1, Hongo, Bunkyo-ku, Tokyo 113, Japan\\
$^2$Institute for Solid State Physics, University of Tokyo 
7-22-1, Roppongi, Minato-ku, Tokyo 106 Japan
}
\date{\today}
\maketitle
\begin{abstract}
We investigate the effects of randomness 
in a strongly correlated electron model
in one-dimension at half-filling.
The ground state correlation functions are 
exactly written by products of 3$\times$3 transfer matrices
and are evaluated numerically.
The correlation lengths depend on randomness
when the interaction is effectively weak.
On the contrary, they are completely insensitive to randomness 
when the interaction is effectively strong.
\end{abstract}

\narrowtext


\vspace{15mm} 
\newpage

The behavior of electrons in the presence of randomness has 
attracted a lot of attentions as one of the most fundamental problems
in condensed matter physics \cite{review}.
In the absence of interaction,
the scaling theory gives us a criterion 
whether states are localized or not \cite{AALR}.
It was shown rigorously that all the states are localized 
for wide class of models in one dimension \cite{Ishii}.
In two dimensions, it is believed that all the states are localized.
Randomness induces a metal-insulator transition
in three dimensions.
So far, a lot of theoretical and experimental works have been reported.
However, the validity to describe the experiments 
by non-interacting models is an open question,
since Coulomb interaction between electrons is always present.

In one dimension, 
some interacting models without randomness can be solved exactly
by the Bethe ansatz technique or bosonization \cite{BA}
and the properties have been investigated.
However,  approaches of including randomness to such models
seem to be hopeless.
In the presence of randomness and without interaction, 
exact results were obtained on the localization
of eigenstates \cite{Ishii}.
In this way, the individual investigations of models 
with interaction or randomness have been succeeded considerablely.
However, it is extremely hard task
to take into account them simultaneously
and we have no reliable method to analize.
Although a few results by the perturbation method \cite{review}
or by bosonization \cite{FKup} are known,
even the qualitative understandings are far from satisfactory.
Numerical investigations have difficulty
due to the restriction of the system size.
Furthermore, one needs enormous amount of CPU time 
for averaging over samples to obtain enough accuracy. 

In this paper, we study a special model at half-filling
to avoid the numerical difficulties mentioned above
to investigate the effect of randomness
in the strongly correlated electron model.
The lattice structure is shown in Fig.~\ref{figlattice} 
and the Hamiltonian with the open boundary condition is given by
\begin{eqnarray}
H &=& {\cal P}
\sum_{\sigma = \uparrow, \downarrow}
\Big\{ \sum_{i=1}^{L}
\Big[
\Big(- p_{i \sigma}^{\dagger} p_{i \sigma}^{\phantom{\dagger}}
     -t_i^{\phantom{\dagger}} p_{i \sigma}^{\dagger} d_{i
\sigma}^{\phantom{\dagger}}
     -t_i^{\phantom{\dagger}} p_{i+1 \sigma}^{\dagger} d_{i
\sigma}^{\phantom{\dagger}}
 +h.c.\Big)
 +V_i^d n_{i \sigma}^d \ \Big]
 + n_{1 \sigma}^p  + n_{L+1 \sigma}^p  \Big\}
 {\cal P},
\label{eqhamiltonian}
\end{eqnarray}
where a unit cell is labeled by $i$.
Here $p_{i \sigma}$ is the annihilation operator
with spin $\sigma=\uparrow, \downarrow$
at site $i$ with no interaction (a $p$-site)
which have at most two electrons with opposite spins,
$d_{i \sigma}$ is the the annihilation operator
at a site with infinitely large on-site Coulomb repulsion
(a $d$-site) which can have at most one electron,
and $n_{i \sigma}^{\alpha}$ ($\alpha =p$, $d$) 
is the electron number operator.
The projection operator which represents
the infinitely large on-site Coulomb repulsion on $d$-sites is
${\cal P} = \prod_i (1 - n_{i \uparrow}^d n_{i \downarrow}^d)$.
We denote the on-site potentials for $d$-sites by $V^d$'s.
For simplicity we parametrize $t$'s and $V^d$'s
by positive $\lambda$'s as

\begin{eqnarray}
t_i &=& \lambda_i \\
V_i^d &=& - 2 \lambda_i^2 +2.
\label{eqparameter}
\end{eqnarray}
Then the on-site potentials at $p$-sites are set to be 
zero except at the boundaries.
We shall take $\lambda$'s to be independent random variable.
The advantages of the model (\ref{eqhamiltonian})
is that the exact and unique ground state 
(at half-filling) is explicitly written  as

\begin{equation}
\Big\vert \Psi_0 \Big\rangle = 
{\cal P}
\prod_{i=1}^L
\prod_{\sigma = \uparrow, \downarrow}
\left( 
       p_{i \sigma}^{\dagger}
     + p_{i+1 \sigma}^{\dagger}
     + \lambda_i^{\phantom{\dagger}} d_{i \sigma}^{\dagger}
\right)
\Big\vert 0 \Big\rangle.
\label{eqgroudstate}
\end{equation}

Without randomness, namely when $\lambda$'s are uniform,
the exact ground state in a restricted parameter space
was obtained \cite{St93} by following the construction
introduced by Brandt and Giesekus \cite{BG92}.
The correlation functions and the momentum distribution functions
were calculated exactly \cite{BL93,YHHK95}.
The correlation functions are exactly represented by products of 
the 3$\times$3 transfer matrices \cite{YHHK95,Tasaki}.
Therefore, the correlation functions and the correlation lengths
can be numerically obtained for considerably long chains
even for random chains.

When there is neither interaction nor randomness,
the ground state is a band insulator.
With interaction and without randomness,
the ground state is also an insulator
but totally different type due to the existence of a spin gap
\cite{KHKup}.
Then the model enables us to investigate 
the effect of randomness to an insulating state 
(where the insulating behavior is due to the strong correlation).
The extension of the method to other parametrizations of the model
and other models of similar type is straightforward.

We take uniform randomness with width $W$

\begin{eqnarray}
\lambda - \frac{W}{2} \le \lambda_i (= t_i) \le \lambda + \frac{W}{2}.
\end{eqnarray}
The probability density function for $W \le 2 \lambda$ is

\begin{eqnarray}
\rho(x=V_i^d)
&=&
\left \{ \begin{array}{@{\,}ll}
          \frac{1}{ W \sqrt{8(2-x)}}
  & \mbox{for $-2\left(\lambda +\frac{W}{2}\right)^2 +2
           \le x \le -2\left(\lambda -\frac{W}{2}\right)^2 +2$ } \\
          0 
  & \mbox{otherwise} \\
      \end{array} \right.
\label{eqprovabilitypot}
\end{eqnarray}
and for $W \ge 2 \lambda$

\begin{eqnarray}
\rho(x=V_i^d)
&=&
\left \{ \begin{array}{@{\,}ll}
      \frac{1}{W \sqrt{8(2-x)}}
  & \mbox{for $-2\left(\lambda + \frac{W}{2}\right)^2 +2$
           $\le x$ $\le -2\left(\lambda - \frac{W}{2}\right)^2 +2$ } \\
      \frac{1}{W \sqrt{2(2-x)}}
  & \mbox{for $-2\left(\lambda - \frac{W}{2} \right)^2 +2 \le x < 2$ } \\
      0 
  & \mbox{otherwise} \\
  \end{array} \right.
\end{eqnarray}
For all $W$, average is

\begin{eqnarray}
\overline{ V_i^d }= -2 \left(\lambda^2 + \frac{W^2}{12} \right)+2.
\end{eqnarray}
The difference of the on-site potentials between $p$- and $d$-sites
depends both on $\lambda$ and $W$.

Without randomness,
the spin, density, singlet-pair, and $\langle
c_{i \sigma}^{\dagger} c_{j \sigma}^{\phantom{\dagger}} \rangle$
correlation functions decay exponentially
\cite{BL93,YHHK95}.
This suggests the existence of a finite excitation gap 
above the ground state
and it was numerically confirmed \cite{KHKup}.
Of course, on $d$-sites Coulomb interaction is always infinity. 
However, by choosing on-site potential $V^d$,
the model interpolates between the following two limits:
(i) $\lambda \to \infty$.
In this limit $ V^d \to -\infty$
and $\langle n^d \rangle$ $\to$ $1$.
Since each $d$-site is occupied by one electron,
they forbid to have additional electrons. 
In this sense the effective interaction is strong.
(ii) $\lambda$ $\to 0$.
In this limit $ V^d \to 2$.
Since the hopping matrix elements between $p$- and $d$-sites,
$\lambda$, is infinitesimal comparing with $V^d$,
one has $\langle n^d \rangle$ $\to$ $0$.
Thus the effective interaction is weak 
since no electron is in $d$-sites.

The occupation and the correlation functions
are exactly written \cite{YHHK95}

\begin{eqnarray}
\langle n_i^{\alpha} \rangle
&=&
\frac{
\langle \Phi_{G.S.} \vert
n_i^{\alpha}
\vert \Phi_{G.S.} \rangle 
}
{
\langle \Phi_{G.S.} \vert \Phi_{G.S.} \rangle 
}
\nonumber \\
&=&
\frac{
\vec{I}^t 
\left( \prod_{k=1}^{i-1} T_k \right)
N_i^{\alpha}
\left( \prod_{k=i+1}^{L} T_k \right)
 \vec{F}
}
{
\vec{I}^t \left( \prod_{k=1}^{L} T_k \right)  \vec{F}
}
\label{equationnumber} \\
\langle {\cal O}_i^{\alpha} {\cal O}_j^{\beta} \rangle
&=&  
\frac{
\langle \Phi_{G.S.} \vert
 {\cal O}_i^{\alpha}
 {\cal O}_j^{\beta}
\vert \Phi_{G.S.} \rangle 
}
{
\langle \Phi_{G.S.} \vert \Phi_{G.S.} \rangle 
}
-\langle  {\cal O}_i^{\alpha} \rangle \langle  {\cal O}_j^{\beta} \rangle
\nonumber \\
&=&
\frac{
\vec{I}^t 
\left( \prod_{k=1}^{i-1} T_k \right)
 O_i^{\alpha}
\left( \prod_{k=i+1}^{j-1} M_k \right)
 O_j^{\beta}
\left( \prod_{k=j+1}^{L} T_k \right)
 \vec{F}
}
{
\vec{I}^t \left( \prod_{k=1}^{L} T_k \right)  \vec{F}
}
-\langle  {\cal O}_i^{\alpha} \rangle \langle  {\cal O}_j^{\beta} \rangle,
\label{eqation2point}
\end{eqnarray}
where 

\begin{eqnarray}
\vec{I}
=\left[
 \begin{array}{c}
  1 \\
  2 \\
  1 \\
 \end{array}
 \right], \ \
\vec{F}
=\left[
 \begin{array}{c}
  1 \\
  0 \\
  0 \\
 \end{array}
 \right].
\end{eqnarray}
Here ${\cal O}^{\alpha}$'s
are the number, spin, creation (anihilation) of singlet-pair,
or creation (anihilation) operators, 
and $\alpha$, $\beta$ = $p$ or $d$.
The matrices $T$'s, $M$'s, and $O$'s 
are the corresponding transfer matrices is given by 

\begin{eqnarray}
& & \ \ \ \ T_n
=
\left[
 \begin{array}{ccc}
     2 \lambda_i^2 + 1
  &    \lambda_i^2 + 1
  &    1               \\
     2 \lambda_i^2
  &  2 \lambda_i^2 + 1
  &  2                 \\
     0
  &    \lambda_i^2
  &  1
 \end{array}
\right],\\ 
& &
\left \{ \begin{array}{@{\,}lll}
M_k = 
-\left[
 \begin{array}{cc}
     \lambda_i^2 + 1
  &  \lambda_i^2   \\
     1
  &  1             \\
 \end{array}
\right]
\\
O_i^p=
\left[
 \begin{array}{ccc}
     \lambda_i^2
  &  \lambda_i^2 + \frac{1}{2}
  &  1                 \\
     0
  &  \lambda_i^2
  &  1
 \end{array}
\right]
&
O_j^p=
-\left[
 \begin{array}{cc}
     \lambda_i^2 + \lambda_i^4
  &  \lambda_i^2  \\
     \lambda_i^4
  &  \lambda_i^2  \\
     0
  &  0
 \end{array}
\right]
        & \mbox{ for the correlation functions 
$\langle c_{i \sigma}^{\dagger}
 c_{j \sigma}^{\phantom{\dagger}} \rangle$} \\
O_i^d=
-\left[
 \begin{array}{ccc}
     \lambda_i
  &  \frac{1}{2} \lambda_i
  &  0                    \\
     0
  &  \frac{1}{2} \lambda_i
  &  0
 \end{array}
\right]
&
O_j^d=
-\left[
 \begin{array}{cc}
     \lambda_i
  &  \lambda_i \\
     \lambda_i
  &  2 \lambda_i \\
     0
  &  \lambda_i
 \end{array}
\right]
 \end{array}
\right.\\
& &\left \{ \begin{array}{@{\,}lll}
M_i=1,
\ \ 
\\
O_i^p
=\left[
 \begin{array}{ccc}
     0
  & -\lambda_i^2
  & -1
 \end{array}
\right]
&
O_j^p=
\left[
 \begin{array}{c}
     \frac{1}{2} \\
     0           \\
    0
 \end{array}
\right]
        & \mbox{for the spin correlation functions} \\
O_i^d=
\left[
 \begin{array}{ccc}
    -\lambda_i^2
  &  0
  &  0
 \end{array}
\right]
&
O_j^d=
\left[
 \begin{array}{c}
     \frac{1}{2} \lambda_i^2 \\
                 \lambda_i^2 \\
    \frac{1}{2} \lambda_i^2
 \end{array}
\right]
      \end{array} \right. \\
& &\left \{ \begin{array}{@{\,}ll}
M_k = T_k,
\ \ 
\\
O_i^p=O_j^p=
\left[
 \begin{array}{ccc}
     \lambda_i^2 + 1
  &  \frac{1}{2}
  &  0                 \\
     2 \lambda_i^2
  &  \lambda_i^2 + 1
  &  1                 \\
     0
  &    \lambda_i^2
  &  0
 \end{array}
\right]
        & \mbox{for the density correlation functions} \\
O_i^d=O_j^d=
\left[
 \begin{array}{ccc}
     \lambda_i^2
  &  \lambda_i^2
  &  0                 \\
     \frac{1}{2} \lambda_i^2
  &  \lambda_i^2
  &  \frac{1}{2} \lambda_i^2 \\
     0
  &  0
  &  0
 \end{array}
\right]
      \end{array} \right. \\
& &\left \{ \begin{array}{@{\,}lll}
M_i=1,
\ \ 
\\
O_i^p
=\left[
 \begin{array}{ccc}
     0
  &  \lambda_i^2
  &  1
 \end{array}
\right]
&
O_j^p=
\left[
 \begin{array}{c}
     1    \\
     0    \\
     0
 \end{array}
\right]
        & \mbox{for the singlet-pair correlation functions} \\
O_i^d=
\left[
 \begin{array}{ccc}
     0  
  &  \lambda_i
  &  0
 \end{array}
\right]
&
O_j^d=
\left[
 \begin{array}{c}
    2 \lambda_i \\
    2 \lambda_i \\
    0
 \end{array}
\right]
      \end{array} \right.
\end{eqnarray}
The correlation lengths of the correlation functions
between $p$-sites and between $d$-sites are the same
up to order $O(1/L)$, since only the matrices at $i$ and $j$ sites
are different in the representation (\ref{eqation2point}).
Due to the same reason, the correlation lengths of the 
spin and the singlet-pair correlation functions are the same
up to order $O(1/L)$.
For a fixed set $\{ \lambda_i \}$, we numerically evaluate the quantities 
\begin{eqnarray}
\overline{\langle n_i^{\alpha} \rangle}
&\equiv&
\frac{1}{N}
\sum_{i=L_B}^{L_B+N}
\langle n_i \rangle
\nonumber \\
\overline{\langle {\cal O}_i^{\alpha} {\cal O}_j^{\beta} \rangle}_{m=j-i} 
&\equiv&
\frac{1}{N}
\sum_{i=L_B}^{L_B+N}
\langle {\cal O}_i^{\alpha} {\cal O}_{i+m}^{\beta} \rangle,
\label{eqaveragecorr}
\end{eqnarray}
where $N$ is the number of sites which are used
for the averaging in a sample
and $L_B$ is the number of sites which is ignored
to exclude contributions form the boundary.
We choose $L=10000$, $L_B=2500$, and $N=5000$.
The occupations of $d$-sites are shown
in Fig.~\ref{figoccupation}.
The sizes of the error bars are smaller than those of the plotted points.
Note that $\langle n_i^p \rangle$ $=2-\langle n_i^d \rangle$,
since the system is half-filled.
We confirmed that the correlation functions decay exponentially.
The correlation lengths are given from
$\overline{\langle {\cal O}_i^{\alpha} {\cal O}_j^{\beta} \rangle}_{m=j-i}$
$\propto$ $\exp{[-m/\xi_{O}]}$,
where $O$ $=S$ for the spin
and
$O$ $=c$ for the correlation function 
$\langle c_{i \sigma}^{\dagger} c_{j \sigma}^{\phantom{\dagger}} \rangle$.
The correlation lengths are estimated by
least squre fit for the values 
$\log\left[\overline{
\langle {\cal O}_i^{\alpha} {\cal O}_j^{\beta} \rangle
}_{m=j-i}\right]$.
The estimates of $\xi_c$ and $\xi_S$ are shown in 
Figs.~\ref{figgreencorrlength} and \ref{figspincorrlength},
respectively. 
The size of the error bars is smaller than that of the plotted points.

The behavior of the correlation lengths
depends on the occupation of $d$-sites,
namely the effective interaction.
For the parameter regime $\lambda \ll 1$
where the effective interaction is weak,
the correlation lengths become short as $W$ increases.
This behavior seems to be similar to the non-interacting cases.
For $\lambda \gg 1$ where the effective interaction is strong,
the correlation lengths are independent of
the strength of the randomness.
We obtained similar behaviors 
for the density and the singlet-pair correlation functions.

For the non-interacting cases, 
spin degree of freedom has nothing to do 
with the properties of the systems.
 For the interacting cases, on the other hand,  spin degree of freedom
plays an important 
role and
the effects of randomness are likely to be different 
from those for the non-interacting cases.
The ground state (\ref{eqgroudstate}) is given
by  superpositions of spin singlet states.
Within the analysis of this model,
the results suggest that the states 
where the effective interaction is strong
have a tendency to have local nature
and the overlappings do not  contribute much
to the expectation values of the correlations.
Thus their properties are stable against randomness.

The authors are grateful to H. Tasaki and 
Y. Hatsugai for useful discussions and comments.  
One of the authors (M.Y.) is supported
by the JSPS Research Fellowships for 
Young Scientists.

\begin{figure}
\caption{
The lattice structure.
An open circle denotes a $p$-site (with no interaction)
and a solid circle denotes a $d$-site (with infinitely large 
on-site Coulomb repulsion).
A line represents hopping of electrons.
\label{figlattice} 
}
\end{figure}

\begin{figure}
\caption{
The estimates of $\overline{\langle n_i^d \rangle}$
as functions of $W$ for $\lambda$ $=0.1$, $0.2$, $0.5$, $1$, and $10$.
\hspace{40mm}
\label{figoccupation}
}
\end{figure}

\begin{figure}
\caption{
The estimates of $\xi_c$
as functions of $W$ for $\lambda$ $=0.1$, $0.2$, $0.5$, $1$, and $10$.
\hspace{40mm}
\label{figgreencorrlength}
}
\end{figure}

\begin{figure}
\caption{
The estimates of $\xi_S$
as functions of $W$ for $\lambda$ $=0.1$, $0.2$, $0.5$, $1$, and $10$.
\hspace{40mm}
\label{figspincorrlength}
}
\end{figure}


\begin{references}
\bibitem[*]
{}Electronic address: yamanaka@appi.t.u-tokyo.ac.jp
\bibitem[**]
{}Electronic address: kohmoto@issp.u-tokyo.ac.jp
\bibitem{review}
For reviews, see
P.A.~Lee and T.V.~Ramakrishnan,
Rev. Mod. Phys. {\bf 57}, 287 (1985);
D.~Belitz and T.R.~Kirkpatrick, 
Rev. Mod. Phys. {\bf 66}, 261 (1994).
\bibitem{AALR}
E.~Abrahams, P.W.~Anderson, D.C.~Licciardello, and T.V.~Ramakrishnan,
Phys. Rev. Lett. {\bf 42}, 673 (1979).
\bibitem{Ishii}
For a review, see
K.~Ishii, Prog. Theo. Phys. Suppl. No.~53, 77 (1973).
\bibitem{BA}
See, for example, {\it The Many-Body Problem}, edited by D.C.~Mattis
(World Scientific, Singapole, 1993).
\bibitem{FKup}
S.~Fujimoto and N.~Kawakami, Phys. Rev. B{\bf 54}, 24640 (1996).
\bibitem{St93}
R.~Strack, Phys. Rev. Lett. {\bf 70}, 833 (1993).
\bibitem{BG92}
U.~Brandt and A.~Giesekus, Phys. Rev. Lett. {\bf 68}, 2648 (1992).
\bibitem{BL93}
P.-A.~Bares and P.A.~Lee, Phys. Rev. B{\bf 49}, 8882 (1993).
\bibitem{YHHK95}
M.~Yamanaka, S.~Honjo, Y.~Hatsugai, and M.~Kohmoto,
J. Stat. Phys. {\bf 84}, 1133 (1996).
\bibitem{Tasaki}
H.~Tasaki, Phys. Rev. Lett. {\bf 70}, 3303 (1993);
Phys. Rev. B{\bf 49}, 7763 (1993).
\bibitem{KHKup}
K.~Kimura, Y.~Hatsugai, and M.~Kohmoto, unpublished.
\end{references}
\end{document}